# A Robust Password-Based Multi-Server Authentication Scheme

**Vorugunti Chandra Sekhar**
Dhirubhai Ambani Institute of
Information and Communication
Technology.
Gandhi nagar, Gujarat, India
sekhar.daiict@gmail.com

**Mrudula Sarvabhatla**
S.V University
Tirupathi, India
mrudula.s911@gmail.com

## ABSTRACT
In 2013, Tsai et al. cryptanalyzed Yeh et al. scheme and shown that Yeh et al., scheme is vulnerable to various cryptographic attacks and proposed an improved scheme. In this poster we will show that Tsai et al., scheme is also vulnerable to undetectable online password guessing attack, on success of the attack, the adversary can perform all major cryptographic attacks. As apart of our contribution, we have proposed an improved scheme which overcomes the defects in Tsai et al. and Yeh et al. schemes.

## Categories and Subject Descriptors
D.4.6, K.6.5

## General Terms
Security

## 1. INTRODUCTION
In Tsai et al., scheme the malicious server $S_j$ can perform undetectable online password guessing attack as follows. Guess PW of $U_i$ as PW1, a11, compute $V_{i1} = h(PW1)$ and sends $\{ID_i, E_{V_{i1}}(g^{a11}, r_{11})\}$ to RC. On receiving the login request the RC retrieves $V_i$ from $R_i \oplus h(ID_i\|x)$ and decrypts the request to get $g^{a11}$, $r_{11}$. RC computes $E_{V_i}(g^{c_{11}})$ and sends $\{ID_i, E_{V_i}(g^{c_{11}})\}$ to $S_j$. On receiving the login reply $S_j$ decrypts using $V_{i1}$ to get $g^{c_{11}}$ and computes $K_{11} = g^{a11.c11}$ and sends the message $\{ID_i, SID_j, E_{K11}(ID_i, SID_j, r_{11}), E_{Vj}(g^{b11}, H(E_{K11}(ID_i, SID_j, r_{11})), ID_i, SID_j, r_{21})\}$ to RC. Now RC performs following computations: Decrypt:: $D_{Vj}(g^{b11}, H(E_{K11}(ID_i, SID_j, r_{11})), ID_i, SID_j, r_{21}))$ to get $E_{K11}(ID_i, SID_j, r_{11})$ and computes 'R' = $H(E_{K11}(ID_i, SID_j, r_{11}))$ and compares 'R' with the received value. If both are equal, RC proceeds further to compute $K11 = g^{a11.c11}$ and decrypts $D_{K11}(ID_i, SID_j, r_{11})$ to get $\{ID_i, SID_j, r_{11}\}$ and compares $r_{11}$ received in login request with $r_{11}$ received. If both are equal means, the original password $V_i$ and password guessed $V_{i1}$ are same else the malicious server can repeat the same process. In our proposed scheme RC stores a variable $k_i$ which is known only to $U_i$ and RC for each user $U_i$. Hence, in our scheme without increase on computation and communications cost, we have restricted the malicious server to perform undetectable online password guessing attack, and other cryptographic attacks.

## 2. TSAI et al., PASSWORD –BASED MULTI SERVER AUTHENTICATION SCHEME

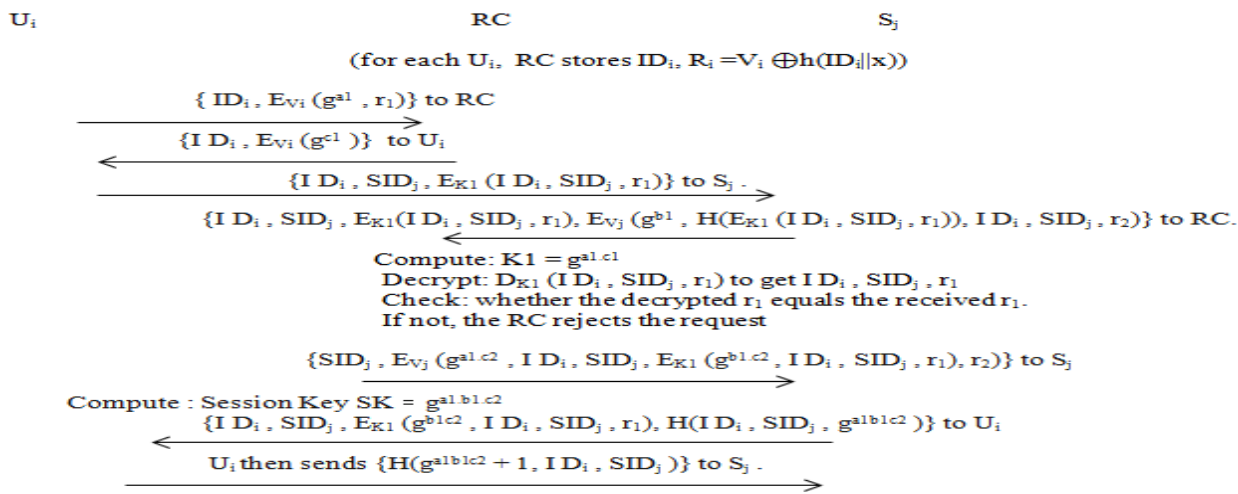

**Figure 1. Graphical View of Tsai et al. scheme.**

## 3. OFFLINE PASSWORD BASED GUESSING ATTACK ON TSAI et al., SCHEME.

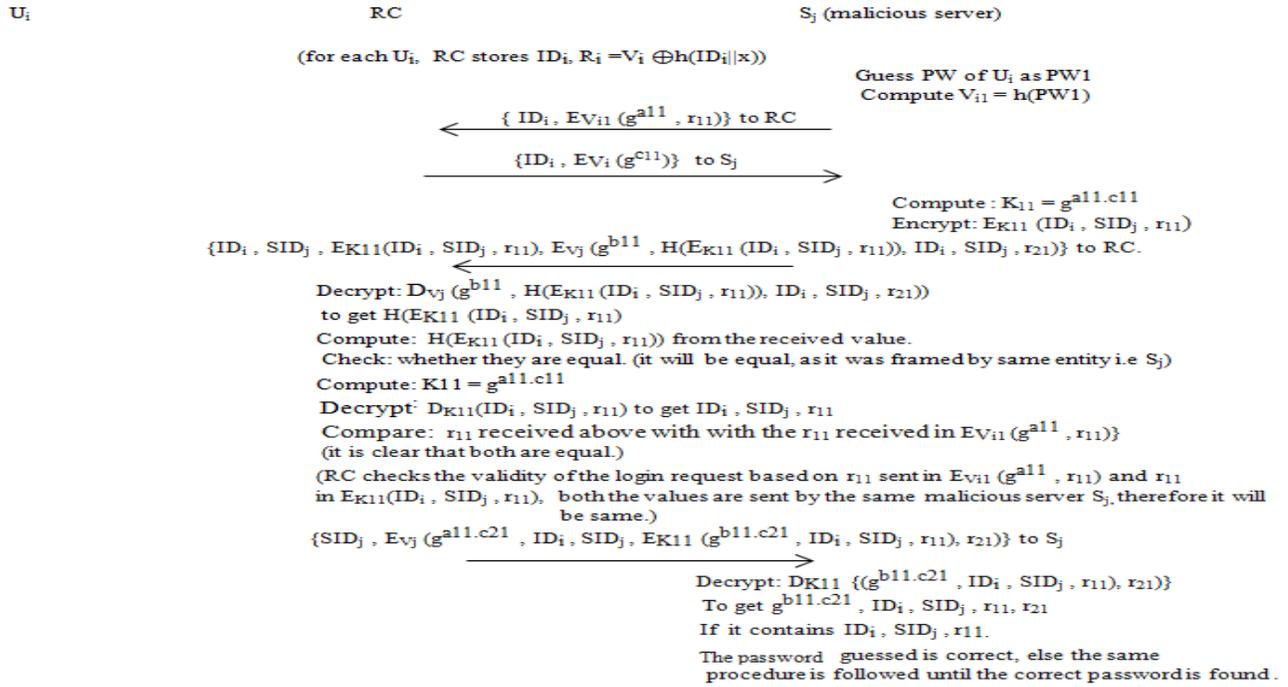

**Figure 2. Offline password guessing attack on TSai et al., scheme.**

## 4. OUR PROPOSED SCHEME

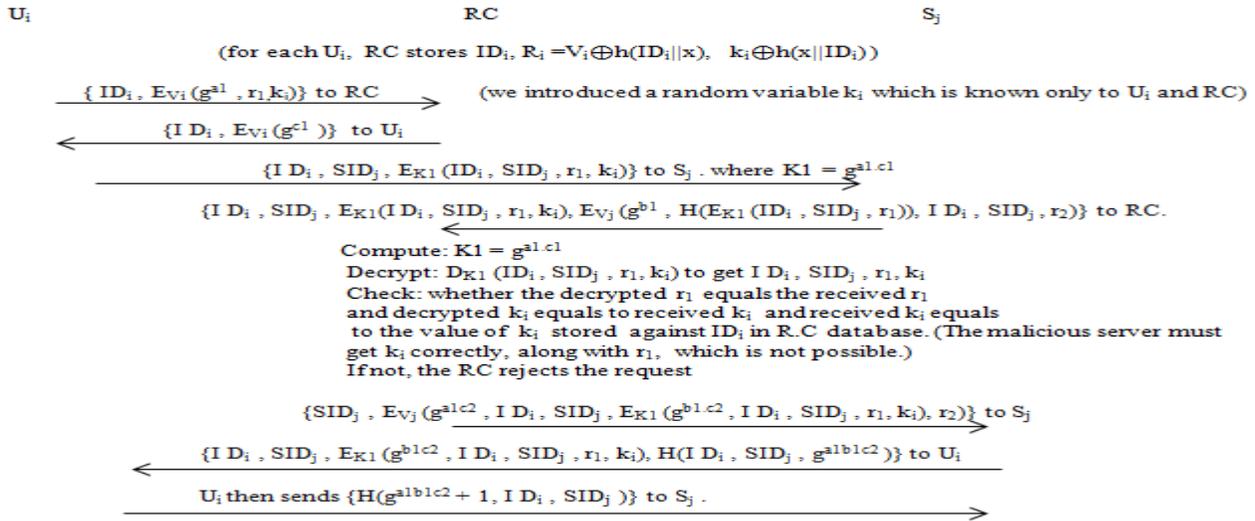

**Figure 3. Our Proposed Scheme.**

## 5. REFERENCES
[1] Lee, J. S., Chang, Y. F., and Chang, C. C. 2008. A novel authentication protocol for multi-server architecture without smart cards. International Journal of Innovative Computing, Information and *Control, 4*(6), 1357–1364.

[2] Yeh, K. H., and Lo, N. W. (2010). A novel remote user authentication scheme for multi-server environment without using smart cards. International Journal of Innovative Computing Information *and Control, 6*(8), 3467–3478.

[3] Tsai, J.L . Lo, N.W  and  W, T.C. (2013).  A New Password-Based Multi-server Authentication Scheme Robust to Password Guessing Attacks. Springer: Journal of Wireless Personal Communications, August 2013, Volume 71, Issue 3, pp 1977-1988.